\documentclass[aps,rmp,twocolumn,eqsecnum,amssymb]{revtex4}
\usepackage{graphicx}
\begin{document}
\title{Adventures in de~Sitter space}
\author{Raphael Bousso}
\affiliation{Institute for Theoretical Physics, 
University of California, Santa Barbara, California 93106, U.S.A.}
\email{bousso@itp.ucsb.edu}
\begin{abstract}

This is my contribution to the Festschrift honoring Stephen
Hawking on his 60th birthday.  Twenty-five years ago, Gibbons and Hawking
laid out the semi-classical properties of de~Sitter space.  After a
summary of their main results, I discuss some further quantum aspects
that have since been understood.  The largest de~Sitter black hole
displays an intriguing pattern of instabilities, which can render the
boundary structure arbitrarily complicated.  I review entropy bounds
specific to de~Sitter space and outline a few of the strategies and
problems in the search for a full quantum theory of the spacetime.
\end{abstract}
\maketitle
\tableofcontents

\section{Introduction}
\label{sec-intro}

It is a pleasure to help celebrate Stephen Hawking's 60th birthday
(not least because Stephen knows how to party).  I am grateful for the
good fortune I had in working with him, and for the physics I learned
from him as his student from 1997 to 2000.  But what I am most
thankful for is the homework.  Stephen's discoveries amount to a
formidable problem set.  I'm afraid we're late turning it in.  Without
question, it will keep us happily occupied for years to come.

Stephen and I have published five journal articles together (Bousso
and Hawking, 1995, 1996b, 1997c, 1998b, 1999a), as well as a number of
proceedings (Bousso and Hawking, 1996a,c, 1997a,b, 1999b).  Most of
these papers, and much of my subsequent work, are concerned in one way
or another with aspects of quantum gravity in de~Sitter space.  In my
contribution to this Festschrift I should therefore like to survey of
some of the results, problems, and speculations surrounding this topic.

Stephen's contributions to black hole physics and cosmology find a
synthesis in his semi-classical treatment of de~Sitter space.  Gibbons
and Hawking (1977) demonstrated that the de~Sitter horizon, like a
black hole, is endowed with entropy and temperature.  Thus, the
quantum properties of black holes will extend to the universe as a
whole, if the vacuum energy is positive.

More than ever, the implications of this work are under active
investigation.  The Bekenstein-Hawking entropy of the spacetime, in
particular, has been taken either as a starting point, or
alternatively, as a crucial test, of various approaches to a full
quantum gravity theory for de~Sitter space.

A brief description of the de~Sitter geometry is given in
Sec.~\ref{sec-classical}.  In Sec.~\ref{sec-bh}, I summarize the
concepts of black hole entropy and the generalized second law of
thermodynamics (Bekenstein, 1972, 1973, 1974), as well as Hawking's
(1974, 1975) results on black hole temperature and radiation.  This
will provide a context for a review of the main conclusions of Gibbons
and Hawking (1977), in Sec.~\ref{sec-ds}.

The generalized second law has been used to infer universal bounds on
the entropy of matter systems.  In Sec.~\ref{sec-bekbound}, I describe
how the Bekenstein (1981) bound, $S\leq 2\pi ER$, is obtained from a
gedankenexperiment involving a black hole.  When this kind of argument
is extended to the de~Sitter horizon, one obtains analogous bounds in
various limits (Schiffer, 1992; Bousso, 2001).  In
Sec.~\ref{sec-dbound}, I discuss one of these bounds, the D-bound,
whose good agreement with the Bekenstein bound is non-trivial.

Roughly speaking, the Bekenstein-Hawking entropy of empty de~Sitter
space is the largest entropy attainable in any asymptotically
de~Sitter spacetime (Fischler, 2000a,b; Banks, 2000).  In
Sec.~\ref{sec-absolute}, I make this absolute entropy bound more
precise.  One must distinguish between spacetimes that are de~Sitter
in the past, in the future, or both.  Moreover, it is natural to
include at least some spacetimes with positive cosmological constant
but no asymptotic de~Sitter region.  The more carefully one
characterizes the spacetime {\em portions\/} whose entropy need be
considered, the broader the {\em class\/} of spacetimes obeying the
bound.  It is argued that only the entropy contained in causal
diamonds is observable (Bousso, 2000).  However, even with this
restriction, one can find some $\Lambda>0$ universes with unbounded
entropy (Bousso, DeWolfe, and Myers, 2002).

The formulation of a quantum gravity theory describing de~Sitter space
remains an open problem.  In Sec.~\ref{sec-qg}, I outline some of the
strategies that can be adopted and the difficulties they face.  In
particular, asymptotic regions, which are conventionally used to
define observables, are not globally accessible to any de~Sitter
observer; quantum mechanical obstructions may be even more severe.  I
also discuss the possibility that a new class of theories, with a
manifestly finite number of states, may play a role in the description
of de~Sitter space.

Sec.~\ref{sec-nariai} is most closely related to my joint work with
Stephen.  It discusses the largest possible black hole in de~Sitter
space.  This spacetime, the Nariai solution, exhibits a remarkable set
of instabilities.  Small perturbations can lead to an infinite variety
of global structures, including the fragmentation of the spatial
geometry into disconnected de~Sitter universes (Bousso, 1998).  
I place these results in the context of present approaches to
de~Sitter space.

This article is by no means an attempt to review the subject.  It
touches upon a small portion of the literature, whose selection is
biased by my current interests and by some of my own adventures in
de~Sitter space.  Many of these were shared with Stephen and with
other collaborators: Andrew Chamblin, Oliver DeWolfe, Andrei Linde,
Alex Maloney, Rob Myers, Jens Niemeyer, Joe Polchinski, and Andy
Strominger.  I would like to thank them, and I emphasize that the
shortcomings of this article, for which I apologize, are mine.

Planck units are used throughout.  For definiteness, the number of
spacetime dimensions is taken to be $D=4$ unless noted otherwise.  The
discussion generalizes trivially to higher dimensions except where
special cases are pointed out.  For a review of de~Sitter space, see,
e.g., Spradlin, Strominger, and Volovich (2001).  Extensive lists of
references are also given by Balasubramanian, de~Boer, and Minic
(2001); Spradlin and Volovich (2001).

\section{de Sitter space}
\label{sec-classical}

This section summarizes a number of the classical properties of
de~Sitter space that are used below.  A more extensive discussion is
found in Hawking and Ellis (1973).  

\begin{figure}\includegraphics[width=7cm]{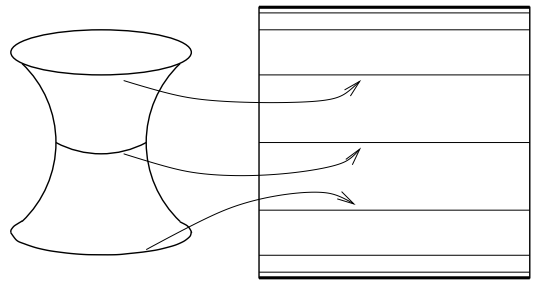}
\caption{de Sitter space as a hyperboloid.  Time goes up.---Right: 
Penrose diagram.  Horizontal lines represent three-spheres.}
\label{fig-closed}
\end{figure}
de~Sitter space is the maximally symmetric solution of the vacuum
Einstein equations with a positive cosmological constant, $\Lambda$.
It is positively curved with characteristic length
\begin{equation}
\ell  = \sqrt{\frac{3}{\Lambda}}.
\label{eq-ell}
\end{equation}
Globally, de~Sitter space can be written as a closed FRW universe:
\begin{equation}
{ds^2\over\ell^2} = - d\tau^2 + \cosh^2 \tau\, d\Omega_3^2
\end{equation}
The spacelike slices are three-spheres.  The space-time can be
visualized as a hyperboloid (Fig.~\ref{fig-closed}).  The smallest
$S^3$ is at the throat of the hyperboloid, at $\tau=0$.  For $\tau>0$,
the three-spheres expand exponentially without bound.  The time
evolution is symmetric about $\tau=0$, so three-spheres in the past
are arbitrarily large and contracting.

The Penrose diagram of de~Sitter space is a square
(Fig.~\ref{fig-closed}).  The spatial three-spheres are horizontal
lines.  As usual, every point represents a two-sphere, except for the
points on the left and right edge of the square, which represent the
poles of the three-sphere.  The top and bottom edge are future and
past infinity, ${\cal I}^+$ and ${\cal I}^-$, where all spheres become
arbitrarily large.

\begin{figure}\includegraphics[width=3.5cm]{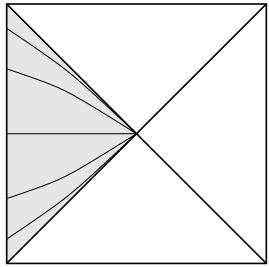}
\caption{Past 
and future event horizon (diagonal lines).  The static slicing
covers the interior of the cosmological horizon (shaded).}
\label{fig-static}
\end{figure}

In the static coordinate system,
\begin{equation}
{ds^2\over\ell^2} = - V(r)\, dt^2 + \frac{1}{V(r)} dr^2 + r^2
d\Omega_2^2,
\label{eq-dsmet}
\end{equation}
where
\begin{equation}
V(r) = 1 - r^2,
\label{eq-dsmet2}
\end{equation}
it becomes manifest that an observer at $r=0$ is surrounded by a
cosmological horizon at $r=1$.  These coordinates cover only part of
the space-time, namely the interior of a cavity bounded by $r=1$
(Fig.~\ref{fig-static}).  This is precisely the operationally
meaningful portion of de~Sitter space, i.e., the region that can be
probed by a single observer.  The upper and lower triangles contain
exponentially large regions that cannot be observed; in particular,
they contain the conformal boundaries ${\cal I}^+$ and ${\cal I}^-$.

An object held at a fixed distance from the observer is redshifted.
The red-shift, $V(r)^{1/2}$, diverges near the horizon.  If released,
the object will accelerate towards the horizon.  Once it crosses the
horizon, it can no longer be retrieved.  In short, the cosmological
horizon acts like a black hole ``surrounding'' the observer.  Note
that the symmetry of the space-time implies that the location of the
cosmological horizon is observer-dependent.  The black hole analogy
carries over to the semi-classical level; this is discussed further in
Sec.~\ref{sec-ds}.

\section{Entropy and temperature of event horizons}
\label{sec-entropy}

\subsection{Black holes}
\label{sec-bh}

The entropy of an ordinary object is lost to an outside observer when
the object falls into a black hole.  However, the black hole's horizon
area increases in this process.  (Indeed, Hawking, 1971, showed that
it never decreases in any classical process.)  In order to salvage the
second law of thermodynamics, Bekenstein (1972, 1973, 1974) proposed
that a black hole carries an entropy on the order of its horizon area,
$A$, in Planck units.  Moreover, he conjectured a generalized second
law of thermodynamics: the sum of ordinary entropy and horizon entropy
never decreases.

The analogy between the laws of thermodynamics and classical
properties of black hole spacetimes---with the surface gravity,
$\kappa$, playing the role of temperature, and the horizon area, $A$,
mimicking entropy---was soon understood in great detail.  Still,
Bekenstein's proposal met with scepticism (Bardeen, Carter, and
Hawking, 1973), because it appeared to lead to a contradiction.  If
$A$ represented an actual entropy, the temperature ($\sim \kappa$) had
to be a physical effect as well.  But how could black holes radiate?

Using Bogolubov transformation techniques, Hawking (1974, 1975)
demonstrated that black holes emit radiation by a quantum process, at
a temperature
\begin{equation}
T_{\rm hor}=\frac{\kappa}{2\pi}.
\label{eq-tkappa}
\end{equation} 
For a Schwarzschild black hole of mass $M$, $\kappa = (4M)^{-1}$.  Via
the first law of thermodynamics,
\begin{equation}
{1\over T_{\rm hor}} = {\partial S_{\rm hor} \over \partial M},
\label{eq-firstlaw1}
\end{equation}
Hawking's calculation confirmed Bekenstein's entropy formula (up to an
additive constant which can be argued to vanish) and determined its
numerical coefficient:
\begin{equation}
S_{\rm hor} = \frac{A}{4}.
\label{eq-shor}
\end{equation}

\subsection{de~Sitter space}
\label{sec-ds}

In spacetimes which are asymptotically de~Sitter in the future, any
observer is surrounded by an event horizon.  At late times, its area
is given by
\begin{equation}
A_0 = 4\pi \ell^2,
\label{eq-a0}
\end{equation}
where
\begin{equation}
\ell = \sqrt{\frac{3}{\Lambda}}
\end{equation}
is the curvature radius, and $\Lambda$ is the cosmological constant.
Gibbons and Hawking (1977) noted that this horizon possesses surface
gravity $\kappa = 1/\ell$ and satisfies analogues to the classical
laws of black hole mechanics (Bardeen, Carter, and Hawking, 1973).

This suggests that the horizons of black holes and of de~Sitter space
share quantum properties as well.  In analogy to Hawking's (1974)
result, one would expect the horizon to be at a non-zero temperature
according to Eq.~(\ref{eq-tkappa}).  Moreover, ordinary matter entropy
is lost when systems cross the de~Sitter horizon.  In analogy to
Bekenstein's (1972) argument, one would expect that the de~Sitter
horizon must have non-zero entropy according to Eq.~(\ref{eq-shor}).

Using Euclidean techniques, Gibbons and Hawking (1977) demonstrated
that an observer in de~Sitter space does detect thermal radiation at a
temperature
\begin{equation}
T_{\rm dS} = \frac{1}{2\pi\ell},
\label{eq-tgh}
\end{equation}
in agreement with expectation.  The presence of thermal Green
functions had been noticed previously in 1+1 dimensional de~Sitter
space (Figari, Hoegh-Krohn, and Nappi, 1975).

One would like to infer the cosmological horizon entropy from the
Gibbons-Hawking temperature by the first law of thermodynamics,
Eq.~(\ref{eq-firstlaw1}).  The mass of a cosmological horizon is not
defined {\em a priori}.  However, only a mass differential is needed.
Hence, let us express the first law in an alternate form, which refers
to the change in matter energy rather than the change in ``horizon
energy''.\footnote{Gibbons and Hawking (1977) formally assign a
negative mass to the cosmological horizon, but this is a mere
convenience.  The differential treatment of energy-momentum flux
across the horizon makes reference only to the matter stress tensor.}

Leaving de~Sitter space aside for a minute, consider a closed system
consisting of a black hole initially well separated from ordinary
matter.  The total energy is conserved in any process whereby energy
is exchanged between the components.  Hence, $dM = -dE$, where $M$ is
the black hole mass and $E$ is the energy of matter, and
Eq.~(\ref{eq-firstlaw1}) becomes
\begin{equation}
{1\over T_{\rm hor}} = -{\partial S_{\rm hor} \over \partial E}.
\label{eq-firstlaw2}
\end{equation}

In this form, the first law makes no reference to the ``energy of the
horizon''.  It can be adapted to the de~Sitter case, where the matter
energy $M$ is perturbatively defined in terms of the timelike Killing
vector field in the interior of the cosmological horizon.  By studying
black hole solutions in de~Sitter space,\footnote{By finding the zeros
of $V(r)$ in Eq.~(\ref{eq-schds2}) as a function of $E$, one estimates
the horizon area as a function of enclosed energy.  Unlike Gibbons and
Hawking (1977), this argument is quick and dirty.  It assumes that a
small Schwarzschild-de~Sitter black hole with mass parameter $ E$ in
Eq.~(\ref{eq-schds2}), and a perturbative Killing mass $E = \int
d^3x\sqrt{h} T_{\mu\nu}\chi^\mu n^\nu$, have equal cosmological
horizon area after back-reaction is taken into account.  (Here
$\chi=\partial/\partial t$ and $n=\chi/|\chi|$.)} one finds that
\begin{equation}
\left. {\partial A_{\rm hor}\over\partial E}\right|_{A_{\rm hor}=A_0}
 = 8\pi\ell.
\label{eq-k}
\end{equation}
for the derivative of the cosmological horizon area [see also
Eq.~(\ref{eq-dboundsmall}) below].  With the Bekenstein ansatz,
$S_{\rm hor}=\eta A_{\rm hor}$, for the entropy of the cosmological
horizon, the usual coefficient, $\eta=1/4$, follows from
Eqs.~(\ref{eq-tgh})--(\ref{eq-k}).  Up to an additive constant, which
is taken to vanish, Eq.~(\ref{eq-shor}) thus applies both to
cosmological and to black hole horizons.  In particular, the total
entropy of empty de~Sitter space is given by its horizon entropy,
which is
\begin{equation}
S_0 = {A_0\over 4} = {3\pi\over\Lambda}.
\label{eq-s0}
\end{equation}

Thus, Gibbons and Hawking (1977) showed that the de~Sitter horizon is
endowed with the same quantum properties as a black hole horizon: a
temperature and an entropy.  They noted that the de~Sitter horizon,
unlike a black hole horizon, is observer dependent.  They interpreted
their results as an indication that quantum gravity may not admit a
single, objective and complete description of the universe.  Rather,
its laws may have to be formulated with reference to an observer---no
more than one at a time.  These insights foreshadow more radical
assertions of the need for complementary descriptions (Susskind,
Thorlacius, and Uglum, 1993), which eventually arose from
considerations of unitarity in the presence of black holes.

\section{Entropy bounds from horizons}
\label{sec-bounds}

\subsection{Black holes and the Bekenstein bound}
\label{sec-bekbound}

When a matter system falls into a black hole, the matter entropy
disappears.  At the same time, the horizon area (and hence, the black
hole entropy) increases.  Bekenstein's generalized second law {\em
may\/} hold, but only if the horizon entropy increases by enough,
i.e., if
\begin{equation}
\frac{\Delta A_{\rm hor}}{4} \geq S_{\rm matter}.
\end{equation}
Bekenstein (1981) estimated a lower bound on $\Delta A_{\rm hor}$
based on the ``Geroch process'', by which a the system of mass $E$ is
added to the black hole only after first extracting a maximum amount
of work.  This minimizes the increase in the black hole mass, and
hence, in its area.

One finds that $\Delta A_{\rm hor} \leq 8\pi ER$, where $R$ is the
largest\footnote{This is an empirical choice.  Classical analysis of
the Geroch process would suggest that $R$ can be the smallest
dimension, which would lead to contradictions.  However, Unruh (1976)
radiation must be taken into account for very flat systems.  More
generally, the proper quantum treatment of the Geroch process is under
debate (see, for example, Bekenstein, 1999; Wald, 2001; Marolf and
Sorkin, 2002).  Independently of its logical status, there is
empirical evidence that Bekenstein's bound holds for all weakly
gravitating matter systems that can actually be constructed (Schiffer
and Bekenstein, 1989; Wald, 2001).  See Bekenstein (2001) and Wald
(2001) for reviews; further references are also given in Bousso
(2002).}  dimension of the system.  Hence,
\begin{equation}
S_{\rm matter} \leq 2\pi ER.
\label{eq-bekbound}
\end{equation}

Remarkably, this conclusion does not depend on microscopic properties
of matter and thus betrays a fundamental aspect of nature.  However,
the bound applies only to matter systems that can actually be added to black
holes.  In particular, one must assume that gravity is not the
dominant force in the system.

\subsection{de~Sitter space and the D-bound}
\label{sec-dbound}

Analogous arguments can be made for matter systems crossing the
de~Sitter horizon.  One possibility is to study matter systems which
are very small compared to the cosmological horizon (Schiffer, 1992).
In this case the cosmological constant is exploited only to provide a
horizon; its effect on spatial curvature is negligible over the scale
of the system.  One finds agreement with the Bekenstein bound.  The
Unruh-Wald analysis can also be generalized in this limit (Davies,
1984).

A different possibility is to consider systems whose size is
comparable to the horizon radius.  The resulting entropy bound is
called the D-bound.  The cosmological curvature is significant in
large systems, and it is not obvious that the D-bound will agree with
Bekenstein's bound, which applies to systems which perturb flat space
weakly.  With reasonable definitions of mass and ``largest dimension''
of the system, however, one finds precise agreement.  This section
follows Bousso (2001).

Consider a matter system in an asymptotically de~Sitter spacetime,
i.e., a spacetime in which an observer's causal domain agrees well
with empty de~Sitter space at late times.  The total initial entropy
is the sum of the matter entropy, $S_{\rm matter}$, and the horizon
entropy, Eq.~(\ref{eq-shor}).  The total final entropy is $A_0/4$, the
entropy of empty de~Sitter space.  By the generalized second law,
\begin{equation}
S_{\rm matter} \leq {A_0 - A_{\rm hor}(\mbox{initial})\over 4}.
\label{eq-dbound}
\end{equation}

This is the D-bound.  It holds for any matter system that can be
contained in a causal domain of an asymptotically de~Sitter universe;
no assumptions about weak gravity are necessary.  In particular, the
D-bound predicts that the cosmological horizon will have area $A_{\rm
hor} < A_0$ as long as matter is present.  This can be verified
explicitly for many solutions, e.g., for the Schwarzschild-de~Sitter
spacetimes.  For light matter systems within the cosmological horizon,
the D-bound is more stringent than the holographic bound, $S_{\rm
matter} \leq A_{\rm hor} /4$.  This can be seen readily in the limit
of empty de~Sitter space, for which $A_{\rm hor}\to A_0$ and the
D-bound vanishes.

Equation (\ref{eq-dbound}) is not always the most useful form of the
D-bound.  One would like to evaluate the right hand side in terms of
intrinsic characteristics of the matter system.  Let us assume that
the matter system is light, i.e., it does not affect the de~Sitter
horizon much:
\begin{equation}
A_0-A_{\rm hor}(\mbox{initial}) \ll A_0.
\end{equation}

Next, suitable quantities must be defined.  We seek a definition of
``mass'' in de~Sitter space that makes reference only to the causally
accessible region (which prevents us from using the definition of
Abbott and Deser, 1982), but does not assume a fixed background (which
precludes the use of a timelike Killing vector).  A convenient
definition makes use of the cosmological horizon area.

The mass $E$ of a system in asymptotically flat space could be
alternatively written as a ``gravitational radius'' $R_{\rm g}$, the
radius\footnote{This refers to the area radius of a horizon, defined
as the root of the proper horizon area divided by the area of a unit
sphere. $R_{\rm g}$ and $R_{\rm c}$ are physical lengths, unlike the
coordinate $r$ used in Eqs.~(\ref{eq-dsmet}) and (\ref{eq-schds}).} of
the Schwarzschild black hole with the same mass.  In this vein, let us
define the mass of a matter system in de~Sitter space to be the radius
of the particular Schwarzschild-de~Sitter black hole that leads to the
same value of the cosmological horizon area.

By substituting $\partial A_{\rm hor}\to 4 S_{\rm matter}$ and
$\partial E\to R_{\rm g}/2$ in Eq.~(\ref{eq-k}), the D-bound can be
expressed in the form
\begin{equation}
S_{\rm matter} \leq \pi R_{\rm g} R_{\rm c},
\label{eq-dboundsmall}
\end{equation}
That is, the entropy of a spherical system in de~Sitter space cannot
be larger than $\pi$ times the product of its gravitational radius and
the radius of the cosmological horizon, $R_{\rm c}$.

In order to compare this result with the Bekenstein bound, it is
useful to express Eq.~(\ref{eq-bekbound}) in terms of the
gravitational radius $R_{\rm g} = 2E$:
\begin{equation}
S_{\rm matter} \leq \pi R_{\rm g} R.
\end{equation}
The ``largest dimension'', $R$, plays a role comparable to $R_{\rm
c}$, at least in the sense that the horizon size in de~Sitter space
places an upper bound on the extent of the system.  Hence, the two
bounds agree for large dilute systems in de~Sitter space.  This is not
trivial as the spacetime background differs significantly.  For
smaller systems, the Bekenstein bound is more stringent.  Of course,
in the limit of very small systems, one expects the Bekenstein bound
to hold, since the deviations from flat space will be negligible.

Both the Bekenstein bound and the D-bound can be extended to $D>4$
spatial dimensions.  They continue to agree exactly in the limit of
large dilute systems.  However, surprisingly, one finds that black
holes saturate the Bekenstein bound only for $D=4$.

\section{Absolute entropy bounds in spacetimes with $\Lambda>0$}
\label{sec-absolute}

The generalized second law states, loosely speaking, that the final
entropy is the largest entropy.  In any spacetime which asymptotes to
de~Sitter space this implies that the maximal entropy is given by
Eq.~(\ref{eq-s0}).  This statement was used to derive the D-bound in
the previous section.  However, de~Sitter space has two conformal
boundaries, one in the past, and one in the future.  At fixed value of
$\Lambda$, one can demand the presence of both boundaries, only one,
or neither.

Interestingly, there are in fact several different statements of the
type $S\leq S_0$ for $\Lambda>0$ universes, depending on asymptotic
conditions.  With liberal conditions on the asymptotic structure, $S$
must be rather restrictively defined for $S\leq S_0$ to obtain.  With
stringent boundary conditions, $S\leq S_0$ holds more broadly.

All of these bounds are {\em absolute\/} in the sense that they refer
to the maximal entropy that can be contained (or probed) in a
spacetime.  They are a consequence of (but not as general as) the
holographic bound ('t~Hooft, 1993; Susskind, 1995; Bousso, 1999a,b),
which limits entropy in the neighborhood of arbitrarily chosen
codimension two spatial surfaces, relative to the surface area.

\subsection{dS$^+$ and the second law}
\label{sec-dsplus}

Let us begin with the class of universes familiar from the previous
section.  The set $\mathbf{dS^+}(\Lambda)$ is defined to contain all
spacetimes which possess an asymptotic de~Sitter region in the future,
with cosmological constant $\Lambda$.  I will not make specific
assumptions about matter content either here or below, except to
demand that reasonable energy conditions are satisfied, e.g., the
dominant energy condition (Wald, 1984).

At least for those observers who reach the asymptotic region, the
second law argument can be completed, and one may conclude that the
D-bound holds, and moreover, that the entropy of empty de~Sitter space
provides an upper bound on the total matter and horizon entropy at any
prior time.

The reference to an observer is crucial, however.  The $\mathbf{dS^+}$
class contains, for example, a spatially flat
Friedmann-Robertson-Walker universe that starts with a big bang
singularity, is radiation or matter dominated initially, and dominated
by a cosmological constant at late times.  (This may be a good
approximation to our own universe.)  The spatial extent of this
universe is infinite at all times, and the entropy on a global time
slice is clearly unbounded.  However, the cosmological horizon shields
all but a finite portion of the universe from any single observer.

The statement $S\leq S_0$ refers only to the entropy of matter inside
the observer's horizon (that is, matter in the observer's causal
past), plus the horizon entropy.  Otherwise, it would obviously be
violated.  This restriction is implicit in the generalized second law
argument.  The growth of the de~Sitter horizon to its asymptotic value
at late times can only be a response to the matter that actually that
crosses that horizon.  The horizon has no knowledge of the entropy of
matter that was already beyond it to start with.

In a dS$^+$ universe, the generalized second law thus implies the
following result.  {\em Consider an observer whose worldline
approaches ${\cal I}^+$, the future de~Sitter infinity, and let $P$ be
the causal past of the point where the worldline meets ${\cal I}^+$.
Let $\Sigma$ be any (suitably smooth and complete) timelike
hypersurface.  The spatial region $\Sigma'\equiv\Sigma \cap P$
corresponds to a particular instant of time in the observer's causal
past, and the boundary of $\Sigma'$ in $\Sigma$, $\partial \Sigma'$,
is the event horizon at that time.  Let $S_{\rm matter}$ be the
entropy of matter in the region $\Sigma'$, and let $A_{\rm hor}$ be
the area of $\partial \Sigma'$.  Then}
\begin{equation}
S_{\rm matter} + \frac{A_{\rm hor}}{4} \leq S_0.
\label{eq-dspbound}
\end{equation}

Note that this argument applies to any observer who reaches ${\cal
I}^+$, but not to observers who fail to do so, for example because
they fall into a black hole.  With suitable energy conditions
preventing the formation of large regions in the interior of a black
hole, one still expects that the entropy in the causal past of such
observers will be bounded as in Eq.~(\ref{eq-dspbound}).  However, I
will not give a detailed argument.

\subsection{dS$^\pm$ and global entropy}
\label{sec-dsplusminus}

Next, consider a subset of $\mathbf{dS^+}$.  Rather than demanding an
asymptotic de~Sitter region only in the future (${\cal I}^+$), let us
insist on a similar region in the past as well (${\cal I}^-$).  The
spacetimes that possess both infinities define the set
$\mathbf{dS^\pm}$.

The Penrose diagram of empty de~Sitter space, Fig.~\ref{fig-static},
is exactly a square, because a light-ray starting in the infinite past
will barely reach the opposite end of the universe in an infinite
time.  If de~Sitter space is not completely empty, the Penrose diagram
will be deformed.  An example is the big bang universe described
earlier; it corresponds to a diagram whose width exceeds its height.
Following Leblonde, Marolf, and Myers (2002), I will call such
diagrams ``short''.  This shape reflects the property that no complete
Cauchy surface is contained in any observer's past.  Examples of this
type preclude a bound on global entropy in the $\mathbf{dS^+}$ class
of universes.

However, if a universe is in $\mathbf{dS^\pm}$, i.e., if it also has a
past asymptotic de~Sitter region, one can use a theorem of Gao and
Wald (2000) to argue that its Penrose diagram is necessarily ``tall'',
except for the exact vacuum de~Sitter solution.  Hence, the observer's
event horizon will cross the entire diagram.  Its area at late times
will be $A_0$; at some finite earlier time $t_{\rm form}$, its area
will vanish.  Hence, there will be early time ($t\leq t_{\rm form}$)
Cauchy surfaces which are completely contained ($\Sigma=\Sigma'$) in
an observer's causal past $P$.  Simply put, an observer in a dS$^\pm$
universe can see what the {\em whole\/} universe looked like in its
early days.

Cauchy surfaces that precede the formation of the event horizon do not
intersect the horizon and hence have $A_{\rm hor}=0$.  Then
Eq.~(\ref{eq-dspbound}) reduces to $S_{\rm matter}\leq S_0$, where
$S_{\rm matter}$ refers to the global entropy on a sufficiently early
Cauchy surface (Bousso, 2002).  The existence of such surfaces is
guaranteed by the Gao-Wald result for all dS$^\pm$ spacetimes (except
for the exact de~Sitter solution, in which the matter entropy
classically vanishes at all times).\footnote{The result of Gao and
Wald (2000), and hence, our conclusion, relies on a number of
technical requirements.  In particular, the spacetime must be
geodesically complete.  The presence of both infinities is not
sufficient to guarantee geodesic completeness even with reasonable
smoothness requirements, because black holes can form.  However, one
would not expect the geodesic incompleteness due to black hole
singularities to invalidate the above conclusions (Susskind,
Thorlacius, and Uglum, 1993).}

To summmarize, Eq.~(\ref{eq-dspbound}) and the theorem of Gao and Wald
(2000) imply the following statement: {\em Consider a spacetime in
$\mathbf{dS^\pm}$, and let $S_{\rm matter}$ be the total matter
entropy on a sufficiently early Cauchy surface.  Then}
\begin{equation}
S_{\rm matter}\leq S_0.
\label{eq-dspmbound}
\end{equation}

The boundedness of the global entropy may seem a surprising result,
since the early-time Cauchy surfaces have divergent volume.  Hence, an
arbitrary amount of entropy can be placed on them without significant
local back-reaction.  However, with this choice of coordinates, the
spacetime will be collapsing initially.  The characteristic scale it
will reach before it can re-expand is set by the value of the
cosmological constant.  However, if the matter density becomes larger
than the energy density of the cosmological constant during the
collapsing phase, it begins to dominate the evolution, and the
universe will collapse to a big crunch.\footnote{Weakly perturbed
de~Sitter space is protected from singularity theorems (Hawking and
Ellis, 1973; Wald, 1984) only because the strong energy condition is
violated by a positive cosmological constant.  If matter satisfying
the strong energy condition dominates while the universe contracts,
one expects this protection to break down.}  Then there will be no
future infinity, in contradiction to our assumption.

\subsection{General $\Lambda>0$ 
universes and the covariant entropy bound}
\label{sec-lambda}

Beginning with the set $\mathbf{dS^+}(\Lambda)$ in
Sec.~\ref{sec-dsplus}, an absolute bound was derived on the entropy in
an observer's causal past.  Specializing to the
$\mathbf{dS^\pm}(\Lambda)$ subset in Sec.~\ref{sec-dsplusminus}, this
result was see to imply a bound on global entropy.  Let us now explore
the opposite direction and try to generalize to larger set of
spacetimes.  Can the $\mathbf{dS^+}(\Lambda)$ set be augmented
non-trivially while retaining an absolute entropy bound of the type
$S\leq S_0$?

Let us define $\mathbf{all}(\Lambda)$ to be the set of all spacetimes
with positive cosmological constant $\Lambda$.  Let's further assume
that matter satisfies the dominant energy condition and that the
number of species is not exponentially large.  Note that $\Lambda$
must be taken to be the lowest accessible vacuum energy.  (For
example, if there are scalar fields, $\Lambda$ will refer to the
cosmological constant at the minimum of their potential.)

$\mathbf{dS^+}(\Lambda)$ is a proper subset of
$\mathbf{all}(\Lambda)$, that is, there are spacetimes with
$\Lambda>0$ that do not asymptote to de~Sitter space in the future.
Consider, for example, a closed universe which begins with an initial
singularity (a big bang).  If the density of ordinary matter is
sufficiently high, the universe will cease its expansion before the
cosmological constant can begin to dominate the evolution.  The
universe will then recollapse to a big crunch.  Another example is the
time-reversal of our own universe (supposing that it started with a
big bang and is about to be dominated by a cosmological constant, as
some observations suggest).

It is more difficult to obtain an absolute entropy bound in
$\Lambda>0$ spacetimes without a future de~Sitter region, because the
second law is no longer of any use.  Indeed, it turns out that
$\mathbf{all}(\Lambda)$ contains spacetimes with unbounded observable
entropy (Bousso, DeWolfe, and Myers, 2002).  The significance of these
examples will be discussed later.

Nevertheless, it is possible to prove an entropy bound for an
interesting subset of $\mathbf{all}(\Lambda)$ which includes certain
spacetimes without ${\cal I}^+$.  The place of the generalized second
law is taken by the covariant entropy bound (Bousso, 1999a; Fischler
and Susskind, 1998), which states that the entropy on any {\em
light-sheet\/} (a contracting null hypersurface) will not exceed a
quarter of its largest area (see Bousso, 2002, for a review).  We will
now summarize this argument.

In order to prove an entropy bound in this context, one must take
great care to determine which parts of the spacetime are accessible to
an observer.  In order to show Eq.~(\ref{eq-dspbound}) for
$\mathbf{dS^+}$, it sufficed to restrict to the portion of the
spacetime in the observer's past.  However, it is not really enough
for information to be present in our past; for it to have operational
meaning, it has to get to us---or, at the very least, it has to enter
a spacetime region that we can actively probe.  Such a region has the
shape of a {\em causal diamond\/} (Bousso, 2000).
\begin{figure}\includegraphics[width=5cm]{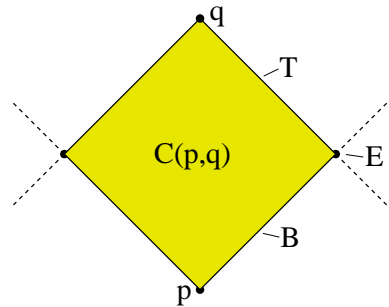}
\caption{Causal 
diamond $C(p,q)$, bounded by top (T) and bottom (B) cone, which
intersect on the edge (E).  A spatial region that fails to fit into
{\em any\/} causal diamond cannot be probed in any experiment.}
\label{fig-cd}
\end{figure}

Given an observer's worldline between two points $p$ and $q$, the
causal diamond $C(p,q)$ is defined by the intersection of the past of
$q$ with the future of $p$ (Fig.~\ref{fig-cd}):
\begin{equation}
C(p,q) \equiv J^-(q) \cap J^+(p).
\end{equation}
In order to determine the largest amount of information available to
any observer in a spacetime ${\cal M}$, it suffices to find an upper
bound on the entropy in the causal diamond $C(p,q)$, for all $p$ and
$q$ in ${\cal M}$.\footnote{Causal diamonds may well play a rather
general role in quantum gravity.  Restriction to regions of the form
$C(p,q)$ may alleviate various problems.  For example, classically, no
more than one point of any spacelike singularity can be contained in a
causal diamond. Thus it is not clear that a quantum treatment of the
vicinity of a spacelike singularity will need to address the
singularity globally.}  To demonstrate an absolute entropy bound for a
whole set of spacetimes, one has to repeat this argument for each
spacetime in the set (or, more efficiently, show that it applies to
all).

The set considered in Bousso (2000) is a subset of
$\mathbf{all}(\Lambda)$, namely the spherically symmetric spacetimes
with cosmological constant $\Lambda$.  Moreover, not all observers
will be considered (i.e., not all $p$ and $q$), but only the observers
which are central (i.e., respect the spherical symmetry).  The entropy
seen by such observers, including horizon entropy if horizons are
present near $C(p,q)$, will not exceed $S_0$.

This is shown as follows.  At the top (the ``future end''), the causal
diamond $C(p,q)$ is bounded by a kind of past light-cone (the boundary
of the past of $q$).  Let us call this the top cone, $T(p,q)$.  By the
second law, the entropy captured on $T(p,q)$ is an upper bound on the
entropy on any other time slice of the causal diamond.  Hence it
suffices to consider $T(p,q)$.

The cone $T(p,q)$ can be thought of as consisting of one light-like
direction and $D-2$ spatial directions.  The latter span
cross-sectional spatial surfaces whose area depends on the position
along the light-like direction.  The focussing theorem implies that
this cross-sectional area takes on precisely one maximal value,
$A_{\rm max}$.  This can be a local maximum along $T(p,q)$ (an
apparent horizon), or it can be reached on the edge, $E$, of the cone,
where $T$ meets a similar cone, $B$, extending from $p$ to the future
(Fig.~\ref{fig-cd}).

In either case, one can show that the surface of maximal area is
either {\em normal\/} (that is, it is neither properly trapped nor
properly anti-trapped), or that its area is smaller than the area of
some normal surface.  Hence, $A_{\rm max}$ must be smaller than the
largest possible normal sphere in any spherically symmetric
$\Lambda>0$ universe.  By Birkhoff's theorem, it must be possible to
match such a normal sphere to a portion of a Schwarzschild-de~Sitter
solution.  One easily verifies that no Schwarzschild-de~Sitter
solution contains any normal spherically symmetric surfaces with area
greater than $4S_0$.  It follows that
\begin{equation}
A_{\rm max}\leq 4S_0.
\end{equation}

Then a rough argument shows immediately that the maximal entropy will
be on the order of $S_0$.  Namely, the surface $A_{\rm max}$ divides
the top cone into one or two portions.  On each portion, the
cross-sectional area decreases in the direction away from $A_{\rm
max}$.  Hence, each portion is a light-sheet of $A_{\rm max}$ and
contains entropy no greater than $A_{\rm max}/4=S_0$.  Therefore,
$S_{\rm matter}\leq 2S_0$.  One also has to take into account the
potential presence of a quasistatic event horizon.  Its area, however,
will not exceed $A_{\rm max}$.  Hence, the total observable entropy,
$S_{\rm matter}+(A_{\rm hor}/ 4)$, will not exceed $3S_0$.

A more detailed analysis (Bousso, 2000) shows that the D-bound,
Eq.~(\ref{eq-dbound}), applies to portions of the top cone in some
cases, yielding tighter inequalities.  Moreover, the presence of a
quasistatic event horizon can be excluded in many cases.  Overall,
these considerations allow a tightening of the above inequality by a
factor of 3.  This leads to the following conclusion.

{\em Observable matter entropy must be contained in a causal diamond.
Let $C(p,q)$ be an arbitrary spherically symmetric causal diamond in a
spacetime with positive cosmological constant $\Lambda$.  Let $S_{\rm
matter}[C(p,q)]$ be the total matter entropy contained in $C(p,q)$.
Define the total observable entropy $S[C(p,q)]$ to be $S_{\rm
matter}[C(p,q)]$ plus the horizon entropy $A_{\rm hor}/4$ of any
quasistatic horizons in the vicinity of the boundary of $C(p,q)$.
Then}
\begin{equation}
S[C(p,q)] \leq S_0.
\label{eq-nbound}
\end{equation}

The assumption of spherical symmetry was necessary in order to apply
Birkhoff's theorem.  This assumption is most relevant for the
spacetimes in $\mathbf{all}(\Lambda) - \mathbf{dS^+}(\Lambda)$, where
other arguments for an absolute entropy bound are lacking.  For
observers which reach ${\cal I}^+$ in a dS$^+$ spacetime, the
bound~\ref{eq-nbound} holds generally.  We expect that
Eq.~(\ref{eq-nbound}) holds for all observers in $\mathbf{dS^+}$, but
this has not been proven.

Yet, the fact that Eq.~(\ref{eq-dspbound}) is independent of spherical
symmetry in $\mathbf{dS^+}$ suggests that its validity in
$\mathbf{all}(\Lambda)$ is more generic than the preceding proof.  It
would be surprising if Eq.~(\ref{eq-nbound}) could be violated by
small deviations from spherical symmetry.  Hence, it was conjectured
in Bousso (2000) that Eq.~(\ref{eq-nbound}), the ``$N$-bound'', holds
generally in $\mathbf{all}(\Lambda)$.

However, the $N$-bound is in fact violated for some spacetimes in
$\mathbf{all}(\Lambda)$ (Bousso, DeWolfe, and Myers, 2002).  The known
counterexamples are topologically different from spherically symmetric
solutions.  Thus, the possibility remains that a set of spacetimes
larger than $\mathbf{dS^+}(\Lambda)$ but smaller than
$\mathbf{all}(\Lambda)$ can be identified for which the $N$-bound
holds generally.  The existence of such a set may be of some
significance for the prospects of quantum gravity theories with a
finite number of states.  This will be discussed at the end of the
next section.

\section{Quantum gravity in de~Sitter space}
\label{sec-qg}

There is evidence that we possess a quantum theory of gravity for
certain asymptotically Anti-de~Sitter (AdS) spacetimes, which are
negatively curved.  String theory is known to admit spacetimes of the
form AdS$_m\times M_n$ where $M_n$ is a suitable compact Euclidean
manifold, for some dimensions $(m,n)$.  In these backgrounds, a
non-perturbative definition of the theory (and thus, of quantum
gravity) has been found in terms of a conformal field theory
(Maldacena, 1998; Gubser, Klebanov, and Polyakov, 1998; Witten, 1998).
String theory (and non-perturbatively, M(atrix)-theory; Banks {\em et
al.}, 1997) also defines S-matrix amplitudes in some asymptotically
flat geometries and in other backgrounds.

Rather than attacking the problem of quantum gravity in complete
generality, these successes suggest that progress can be made by
restricting to suitable classes of spacetimes, characterized by
asymptotic boundaries which are protected from quantum fluctuations.
After addressing flat and negatively curved geometries, a natural step
is to consider a positive cosmological constant next.  This case is of
particular interest because it might include the universe we inhabit
(Riess {\em et al.}, 1998; Perlmutter {\em et al.}, 1999).

However, twenty-five years after the semi-classical results of Gibbons
and Hawking (1977), a full quantum gravity theory for asymptotically
de~Sitter spacetimes is still lacking.  In particular, it has turned
out very difficult to realize de~Sitter space in string theory (for
recent approaches, see, e.g., Hull, 2001; Silverstein, 2001; Gutperle
and Strominger, 2002; and the contribution by Maloney, Silverstein,
and Strominger to this volume).

\subsection{To drop infinity or to keep infinity}
\label{sec-drop}

It is not clear how broad a set of spacetimes would be described by a
``quantum theory of de~Sitter space''.  From the experience with
string theory, one expects that a particular matter content,
presumably compatible with reasonable energy conditions, will arise
from the theory.  But what about asymptotic conditions?  Should both
asymptotic regions be demanded ($\mathbf{dS^\pm}$)?  Will the theory
describe the broader class $\mathbf{dS^+}(\Lambda)$, requiring only
that the spacetime asymptote to de~Sitter space in the future?  Or
should we abandon asymptotic conditions entirely and seek a theory of
$\mathbf{all}(\Lambda)$, spacetimes characterized merely by a
particular positive value of the cosmological constant?

No such confusions arise for asymptotically AdS and flat universes,
because their asymptotic boundaries have only one connected component.
Moreover, in the AdS and flat cases, the presence of the asymptotic
region is not affected by continuous changes to Cauchy data.  For
spacetimes with positive cosmological constant, however, small changes
in the stress tensor at one time may affect the presence of asymptotic
de~Sitter regions in the past or future (Bousso, 2000; Bousso,
DeWolfe, and Myers, 2002).  Of course, it is conceivable that a
non-perturbative quantum theory will preclude variations of Cauchy
data that would classically change the asymptotic structure.  However,
at least from a low-energy perspective, this would seem unnatural.

Classically, one must keep in mind that even in a dS$^+$ universe, not
all observers reach future infinity; some fall into black holes.
Moreover, an observer's causal diamond contains at most one point each
of ${\cal I}^+$ and ${\cal I}^-$, so that physical observables cannot
be defined in the asymptotic regions.  At the quantum level, one
expects all structures to be thermalized by Gibbons-Hawking radiation
within finite time, which prevents observers from reaching ${\cal
I}^+$ altogether.  Together with the difficulty mentioned in the
previous paragraph, this makes it desirable to seek a theory which
ultimately does not require or make use of asymptotic de~Sitter
regions (for example a theory of $\mathbf{all}(\Lambda)$, with
suitable matter restrictions).

However, one has no control over metric fluctuations in the spacetime
interior, and no symmetries can be assumed.  This impedes concrete
progress without some reference to asymptotic regions.  Moreover,
whether or not it will ultimately survive in the formulation, the
structure of the de~Sitter infinities may well provide clues to
properties of a quantum gravity theory.  This has been the subject of
numerous studies, especially in the context of the correspondence
between de~Sitter space and a Euclidean conformal field theory
recently conjectured by Strominger (2001a).\footnote{I will not
attempt to survey the literature on this approach.  Extensive lists of
references may be found in Balasubramanian, de Boer, and Minic (2001),
and in Spradlin and Volovich (2001).  In viewing cosmological
evolution as inverse RG flow, Strominger (2001b) has also outlined a
possible approach to understanding the apparent increase in the number
of available degrees of freedom with time.}

\subsection{A theory with finite-dimensional Hilbert space?}
\label{sec-finite}

Reproduction of the entropy $S_0$ of de~Sitter space provides a key
test for any formulation of quantum gravity.  For dS$^\pm$ spacetimes,
Witten (2001) has argued that a Hilbert space of dimension $e^{S_0}$
might arise from a larger space of states via a non-standard inner
product.  In the context of the dS/CFT correspondence, the entropy of
de~Sitter and Kerr-de~Sitter spacetimes has been numerically
reproduced (e.g., Balasubramanian, de~Boer, and Minic, 2001; Bousso,
Maloney, and Strominger, 2001).  However, this was done by methods
whose justification from the CFT point of view is still incomplete.

Fischler (2000a,b) and Banks (2000) have proposed that the finiteness
of the de~Sitter entropy should be elevated to a defining principle
for the theory.  The bound on observable entropy in all dS$^+$
universes, Eq.~(\ref{eq-dspbound}), implies that a finite number of
states suffices to completely describe all of physics in such
universes.  It would be most economical, therefore, to seek a quantum
gravity theory with Hilbert space of finite dimension $e^{S_0}$.
Conversely, perhaps a positive cosmological constant should be
regarded as nature's way of ensuring that entropies greater than $S_0$
simply cannot occur---an essential cutoff if our Hilbert space is
really finite.

The Banks-Fischler proposal suggests that a positive cosmological
constant should not be regarded as a consequence of complicated
quantum corrections and cancellations.  Rather, $\Lambda>0$
constitutes a direct and fundamental reflection of the size of the
Hilbert space of quantum gravity.  A correpondence between the value
of the cosmological constant and the number ${\cal N}$ of states in
the Hilbert space is thus implied.  Thus, the proposal offers a fresh
perspective on the cosmological constant problem.  (This is called the
``$\Lambda$-$N$ correspondence'', where $N=\log {\cal N}$, in Bousso,
2000.)

It must be stressed that this proposal goes beyond what is necessary
to explain de~Sitter entropy.  It exploits the fact that a
cosmological constant can be regarded as a fixed property of a theory,
rather than a variable parameter associated with a solution.  The
finite entropy of a black hole, by contrast, reflects only those
states (of a larger or infinite Hilbert space) which actually
correspond to the black hole.  As the mass of the black hole is
usually considered a variable parameter, it cannot possibly constrain
the dimension of the full Hilbert space, which ought to be infinite
for asymptotically flat or AdS spacetimes.

The Banks-Fischler proposal asserts that the de~Sitter entropy will
not arise from a subset of states, but represents the complete Hilbert
space of a theory.  In particular, there would be no possibility of
describing de~Sitter by a theory with infinitely many states.  This
would be a remarkable constraint.  For example, a single harmonic
oscillator has an infinite-dimensional Hilbert space, not to speak of
quantum field theory or string theory.

The Banks-Fischler proposal assigns a crucial role to theories with
Hilbert space of finite dimension $\cal N$ for the description of
certain cosmological spacetimes.  The physics of asymptotically flat
or AdS universes (e.g., string theory) would be recovered only in the
limit ${\cal N}\to\infty$.  Conversely, this would explain why
de~Sitter space has not been found in string theory.

For $D>4$, the $\Lambda$-$N$ correspondence suffers from the
shortcoming that the specification of a positive cosmological constant
alone does not guarantee the entropy bound~(\ref{eq-nbound}).  As
discussed at the end of the previous section, explicit counterexamples
are known (Bousso, DeWolfe, and Myers, 2002).  In other words,
$\mathbf{all}(\Lambda)$ contains spacetimes with observable entropy
greater than $S_0$.  Such spacetimes cannot possibly be described by a
theory with only ${\cal N}=e^{S_0}$ states.  Some $\Lambda>0$
spacetimes must be excluded from the ``gravity dual'' of any
finite-${\cal N}$ theory.

Thus, a simple relation between the size of Hilbert space and the
cosmological constant cannot hold unless additional conditions are
specified.  Clearly, the demand of a future asymptotic de~Sitter
region is a sufficient condition.  However, as discussed in
Sec.~\ref{sec-drop}, it is both artificial and operationally
questionable to distinguish spacetimes in $\mathbf{dS^+}$ from at
least some of the closely related spacetimes in
$\mathbf{all}(\Lambda)-\mathbf{dS^+}(\Lambda)$.

The search for a suitable completion of the set
$\mathbf{dS^+}(\Lambda)$, in which Eq.~(\ref{eq-nbound}) would hold,
has not yet succeeded.  Such a completion would give support to the
Banks-Fischler proposal.  It would provide a concrete candidate set of
spacetimes that might be described by quantum gravity theories with
finite ${\cal N}$, if such theories exist.

\subsection{Other questions}
\label{sec-fuzzy}

There are many other open questions about de~Sitter space.  If the
region near ${\cal I}^+$ cannot be observed, then what are the
observables?  Is the evolution of matter fields and horizon in
de~Sitter space unitary or is information lost?  

Hawking (1976) claimed that black holes convert pure states to mixed
states.  The debate continues despite recent results in support of
unitarity (Strominger and Vafa, 1996; Maldacena, 1998).  For a
de~Sitter horizon, it is not clear whether the question is even
well-posed.  One can attempt to extend black hole complementarity
(Susskind, Thorlacius, and Uglum, 1993) to de~Sitter space (Dyson,
Lindesay, and Susskind, 2002) and restrict to a causal diamond region
(Bousso, 2000).  However, asymptotic states cannot be defined, and it
is not clear how unitary evolution would be verified in any
experiment.

In particular, a black hole can evaporate completely; in principle, it
can return information in correlations of the Hawking radiation.  On
the other hand, the cosmological horizon never disappears completely
except in the catastrophic collapse of the entire spacetime.  No more
than a third of the degrees of freedom are available in matter form
(this limit arises from the largest black hole in de~Sitter space).
Thus, if the whole system, consisting of matter and the cosmological
horizon, were in a pure state, the matter subsystem would be unlikely
to contain any information at all (Page, 1993).  Finite observer
lifetimes further complicate this problem.

Yet, whether or not we live in de~Sitter space, many of the above
conceptual problems arise in any attempt at a quantum treatment of
cosmology.  de~Sitter space offers a relatively simple arena for their
investigation.

\section{Instabilities of the Nariai solution}
\label{sec-nariai}

\subsection{Schwarzschild-de~Sitter and Nariai}

Black holes in de~Sitter space cannot be larger than the de~Sitter
horizon.  Small Schwarzschild-de~Sitter black holes are much hotter
than the cosmological horizon, and the geometry in their neighborhood
is a good approximation of a Schwarzschild black hole in flat space.
Their evolution will not differ much from their flat space cousins.

In this section I will discuss quantum aspects of black holes which
are of a size comparable to the cosmological horizon.  These ``large''
Schwarzschild-de~Sitter black holes have no flat space analogue.  They
constitute interesting physical systems in their own right.  The
interplay between the two horizons leads to novel effects.
Instabilities arise which can complicate the global structure of
asymptotically de~Sitter spacetimes.

\begin{figure}\includegraphics[width=7cm]{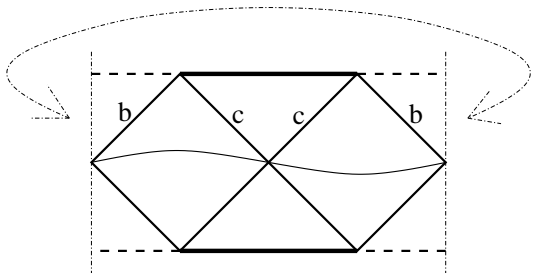}
\caption{Penrose 
diagram of a Schwarzschild-de~Sitter spacetime.  The curved line is a
slice of equal time in the static coordinates; its geometry is a
warped product of $S^1$ and $S^2$.  The $S^2$ directions are
suppressed in this diagram; the $S^1$ arises because the left and
right ends are identified.  The black hole (b) and cosmological (c)
horizons are indicated.  The static coordinates, Eq.~(\ref{eq-schds}),
cover one of the diamond-shaped regions.  The black hole singularity
and the de~Sitter infinity are shown as dashed and bold lines,
respectively.  The Nariai solution has the same Penrose diagram except
for the nature of the boundaries.  The Penrose diagram for a multiple
Schwarzschild-de~Sitter solution is obtained by joining several copies
of this diagram before identifying the ends.}
\label{fig-schds}
\end{figure}

The Schwarzschild-de~Sitter solution is given by the metric
\begin{equation}
{ds^2\over \ell^2}= 
- V(r) dt^2 + {dr^2\over V(r)} + r^2 d\Omega_2^2,
\label{eq-schds}
\end{equation}
where
\begin{equation}
V(r) = 1-{2 E\over r}-r^2.
\label{eq-schds2}
\end{equation}
This metric is static and covers only a portions of the maximally
extended spacetime, as seen in the Penrose diagram of
Fig.~\ref{fig-schds}.

The mass parameter $ E$ grows monotonically with the size of the
black hole.  For $ E=0$ one recovers empty de~Sitter space.  If
$0< E< E_{\rm max}\equiv 3^{-3/2}$, $V(r)$ has two positive zeros,
corresponding to the black hole and the cosmological horizon.  The
fully extended spatial geometry at $t=0$ has topology $S^1 \times
S^2$.  The size of the $S^2$ varies as a function of the coordinate on
the $S^1$; it is minimal on the black hole horizon and maximal on the
cosmological horizon.

The cosmological horizon decreases as $ E$ is increased.  For
$ E= E_{\rm max}$, both horizons are at the same value of $r$, at
$r=3^{-1/2}$.  In the metric~(\ref{eq-schds}) it would appear that
there is no space left in the geometry.  In fact, however, only the
coordinate $r$ becomes degenerate and ceases to be useful.  A proper
limiting procedure (Ginsparg and Perry, 1983) shows that the geometry
of the $t=0$ slice remains perfectly regular as $ E\to  E_{\rm max}$
and becomes the geometry of the Nariai solution.  This space is the
direct product of an $S^1$ and an $S^2$, both with radius
$r=3^{-1/2}$.

\subsection{First-mode instability}

Although the Nariai and Schwarzschild-de~Sitter geometries nearly
agree at an instant of time, they differ markedly in their temporal
evolution.  The $S^1$ factor of the Nariai solution expands
exponentially, forming a 1+1 dimensional de~Sitter spacetime.  The
$S^2$ factor remains constant.  In global coordinates, the Nariai
metric is given by
\begin{equation}
{ds^2\over\ell^2} = \frac{1}{3} \left(
-dT^2 + \cosh^2 T\, dx^2 + d\Omega_2^2 \right).
\label{eq-nariai}
\end{equation}
Unlike all of the Schwarzschild-de~Sitter solutions, the Nariai
spacetime is homogeneous.  It does not possess any singularity, nor
does it possess four-dimensional asymptotic de~Sitter
regions.\footnote{The singularities as well as the asymptotic
boundaries of the Schwarzschild-de~Sitter spacetimes lie in the far
future or past.  They are not places in space.  Hence the different
boundary structure of the Nariai solution does not contradict the
similarity of the spatial metrics at one instant of time [the $T=0$
slice of (\ref{eq-nariai}) and the $ E\to E_{\rm max}$ limit of (the
full extension of) the $t=0$ slice of (\ref{eq-schds})].}

These properties make a classical instability of the Nariai solution
manifest.  Consider a small perturbation of the $T=0$ slice, such that
the two-sphere area is not constant but instead is given by $(\ell^2/
3) (1+\epsilon\cos x)$.  That is, the two-sphere size oscillates once
as a function of the angular variable on the $S^1$.  To leading order,
this will revert the geometry to a nearly maximal
Schwarzschild-de~Sitter metric (Ginsparg and Perry, 1983).  The
two-spheres that are smaller than the Nariai value will collapse to
form the black hole interior.  The larger two-spheres expand
exponentially to generate an asymptotic de~Sitter region.

A nearly maximal Schwarzschild-de~Sitter black hole is classically
stable.  Quantum mechanically one expects Hawking radiation to be
emitted both by the black hole, and by the cosmological horizon that
surrounds it.  In the Nariai solution, the two horizons would be in
equilibrium at a temperature $T={3^{1/2}/( 2\pi\ell)}$ (Bousso and
Hawking, 1996b).  The black hole and the cosmological horizon emit and
receive equal amounts of radiation.

One expects perturbations of the Nariai geometry to disturb this
equilibrium.  This question was studied by Bousso and Hawking (1998b).
The quantum radiation was included in the $s$-wave approximation at
the level of a one-loop effective action. (Different actions were
employed by Nojiri and Odintsov, 1999a,b, 2001).  We found that large
Schwarzschild-de~Sitter black holes are unstable to radiation.

Because of the interplay between the two nearly equal horizons, the
dynamics of the evaporation can be more involved than it is for small
black holes.  For some perturbations, the black hole horizon grows
towards the Nariai value at early times.  However, the shrinking mode
is expected to dominate at late times.

Our analysis was carried out perturbatively about the Nariai solution.
This did not amount to a conclusive argument that the evaporation of
black holes will continue well into the small black hole regime, where
the effect of the cosmological horizon can be neglected and one can be
sure that the evaporation process will in fact complete.  Later this
was demonstrated by a non-linear numerical analysis (Niemeyer and
Bousso, 2000).  Schwarzschild-de~Sitter black holes evaporate, be they
large or small.

\subsection{Higher modes and fragmentation}

So far, I have discussed only a first-mode perturbation.  The Nariai
solution has more in store.  A constant mode
perturbation\footnote{Classically, the perturbations in vacuum must
satisfy certain constraint equations.  In the one-loop model studied
by Bousso and Hawking (1998b) this becomes a constraint on the initial
distribution of the radiation, which can be satisfied for all metric
perturbations at leading order.}  makes the two-sphere area everywhere
smaller, or everywhere larger, than the value $\ell^2/3$.  The former
case is like trying to make a black hole that doesn't fit into
de~Sitter space; the spacetime collapses globally in a big crunch.
The latter case will lead to a solution in which all regions are
locally de~Sitter but the global topology is non-trivial.

\begin{figure}\includegraphics[width=8cm]{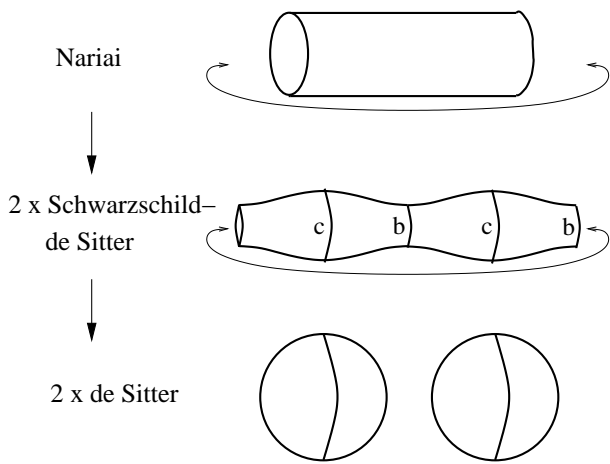}
\caption{Fragmentation.  
Nariai space is a product $S^1 \times S^2$.  The $S^2$ is represented
as an $S^1$ here; the time direction is suppressed in the drawings and
indicated only through the arrows evolving the snapshots of the
spatial geometry.  Upon a higher-mode perturbation (here, $n=2$),
Nariai space can evolve into a sequence of Schwarzschild-de~Sitter
universes.  Black hole (b) and cosmological (c) horizons are
indicated.  When the black holes evaporate, the geometry pinches in
several places, and only disconnected de~Sitter portions remain.}
\label{fig-stepshaw}
\end{figure}

Higher-mode perturbations of the Nariai solution give rise to rather
drastic global effects (Bousso, 1998).  Consider the $n$-th mode
($n>1$), which perturbs the two-sphere radius as $(\ell^2/ 3)
(1+\epsilon\cos nx)$.  If this fluctuation dominates, one finds the
following classical evolution.  The mode is oscillatory at first,
increasing in proper wavelength as the $S^1$ expands.  When the $S^1$
has expanded by a factor $n$, the wavelength becomes larger than the
horizon scale, and the mode grows exponentially.  This marks the
beginning of the formation of $n$ black hole interiors and $n$
asymptotically de~Sitter regions (Fig.~\ref{fig-stepshaw}).

The geometry then resembles a sequence of Schwarzschild-de~Sitter
solutions.  In the ordinary Schwarzschild-de~Sitter solution ($n=1$),
the black hole connects opposite ends of a single asymptotic de~Sitter
region (Fig.~\ref{fig-schds}).  There are two black hole horizons, but
only one black hole interior.  In the solutions obtained for $n>1$,
however, each black hole connects two different de~Sitter regions;
after traversing\footnote{This is a description of spacelike surfaces,
not of causal paths through the spacetime.} $n$ such regions and $n$
black holes, one is back at the first region.

The higher-mode instabilities of the Nariai solution become even more
interesting when Hawking radiation is taken into account (Bousso,
1998).  Perturbatively, each of the black holes is found to evaporate,
much like a large Schwarzschild-de~Sitter black hole in a single
asymptotic region.  At the non-linear level, the evaporation was again
found to continue until the black holes are much smaller than the
cosmological horizons (Niemeyer and Bousso, 2000).  Then nothing can
stabilize them, and one expects that they will disappear altogether.

If $n=1$, the complete evaporation of a Schwarzschild-de~Sitter black
hole can be visualized as a deformation and eventual topological
transition of the $S^1\times S^2$ spatial sections.  The direct
product metric (Nariai) becomes a warped product
(Schwarzschild-de~Sitter), in which the two-sphere size varies along
the $S^1$.  Eventually the $S^2$ vanishes at one point; the $S^1\times
S^2$ geometry pinches off and reverts to an $S^3$ topology (empty
de~Sitter space).  If $n>1$, however, the $S^2$ size decreases at $n$
points on the $S^1$, as all $n$ black holes evaporate.  Hence, the
spatial geometry is pinched in several places, leaving behind $n$
disconnected spatial manifolds, each of topology $S^3$.

Thus, the spacetime fragments.  Notice that this is not an off-shell
Planckian quantum fluctuation of the metric by which some kind of baby
universe is created.  The process is slow and under good
semi-classical control.  Planckian curvatures enter only at the
endpoint of the evaporation, which is usually assumed to correspond to
the disappearance of the black hole.  The fragments can be arbitrarily
large.

\subsection{Global structure, black hole creation, and proliferation}

The instability of the Nariai solution is remarkable in that
arbitrarily small variations in the spatial geometry can lead to an
unlimited variety of causal structures.  The mode number of the
dominant perturbation determines the number of copies of ${\cal I}^+$;
upon inclusion of Hawking radiation, it determines the number of
disconnected components of space at late times.  This can be regarded
as a generalization of an effect discussed in Sec.~\ref{sec-qg} above.
There it was noted that small changes in Cauchy data can determine
whether or not ${\cal I}^+$ is present at all.

In this section, I have described the decay of the Nariai solution,
not the decay of de~Sitter space.  However, it is possible for black
holes to nucleate spontaneously in de~Sitter space.  A number of
arguments lead to this expectation and yield mutually compatible
estimates of the nucleation rate.  For example, black holes can be
assumed to lie in the exponential tail of the thermal radiation
emitted by the cosmological horizon.  However, the gravitational
instanton approach has been most commonly used (Ginsparg and Perry,
1983; Bousso and Hawking, 1995, 1996b, 1999a; Mann and Ross, 1995;
Chao, 1997; Bousso and Chamblin, 1999).  In this picture, the
formation of black holes is regarded as a tunneling event, much like
the Schwinger pair creation of charged particles in a strong electric
field.  One finds at leading order that the rate of black hole
formation is suppressed by the difference between the de~Sitter
entropy and the Schwarzschild-de~Sitter entropy:
\begin{equation}
\Gamma \sim \exp(S_{\rm SdS}-S_0).
\end{equation}

In particular, a Nariai black hole might nucleate, with a rate of
$\exp(-\pi/\Lambda)$.  Thus, the possibility arises that de~Sitter
space proliferates (Bousso, 1998) by the iterated production,
evaporation, and fragmentation of Nariai geometries.  However, it is
not clear whether a global description of the nucleation process,
suggested by the instanton approach, is adequate
(Sec.~\ref{sec-lambda}).  In fact, the repeated nucleation of black
holes at different times, which occurs naturally in a single causal
region from a statistical mechanics point of view, faces global
obstructions in the instanton picture, as the nucleation surfaces
would mutually intersect (Jacobson, unpublished).  For the purpose of
generating an unlimited number of components of ${\cal I}^+$, at
least, those obstructions can be circumvented by considering black
holes of sufficient charge (Bousso, 1999c).

\subsection{Discussion}

The work reported in this section precedes the current surge of
interest in de~Sitter space and accelerating universes.  Many of the
present approaches are guided by the desire to apply string theory, or
at least some of the lessons learned from recent developments in
string theory, to the problem of de~Sitter quantum gravity.  The study
of the Nariai solution might add some useful perspectives to these
endeavors.

It is possible to explore the vacuum structure of string theory by
identifying certain unstable configurations (e.g., the vacuum of
bosonic string theory, or a D-brane/anti-D-brane pair).  The idea is
to study their evolution and try to describe the structure of the
decay product (see, e.g., Sen, 1999).  In this sense, the Nariai
solution may offer a way of circumventing the difficulty of
incorporating de~Sitter space in string theory.  If a Nariai solution
could be constructed, its decay would naturally lead to de~Sitter
space, and perhaps to the more complicated configurations obtained by
the fragmentation process.  Though there is no guarantee that a Nariai
solution can be implemented in string theory, the possibility of this
approach should be noted.

In some of the present approaches to de~Sitter quantum gravity
(Sec.~\ref{sec-qg}), asymptotic boundaries play a central role, and
the presence of a single copy each of ${\cal I}^+$ and ${\cal I}^-$ is
often assumed (e.g., Strominger, 2001a; Witten 2001).  We noted
earlier that the presence of these boundaries is by no means
guaranteed and is sensitive to small variations of Cauchy data.  The
present section has shown that the conformal boundary can become
arbitrarily complicated in de~Sitter-like spacetimes.  There can be an
unlimited number of disconnected components of ${\cal I}^+$.
Moreover, spacetimes with a large variety of different asymptotic
structures can be obtained from small variations of topologically
identical initial conditions.  It is conceivable that this additional
structure might play a role in the formulation of a quantum gravity
theory in de~Sitter space.

\paragraph*{Acknowledgement} 
This work was supported in part by the National Science Foundation
under Grant No.\ PHY99-07949.


\bibliographystyle{myrmp}
\bibliography{all}

\end{document}